\documentclass[aps,preprint,eqsecnum,nofootinbib]{revtex4-1}

\usepackage{epsfig}
\usepackage[usenames, dvipsnames]{color}

\newcommand{\be}{\begin{equation}}
\newcommand{\ee}{\end{equation}}
\newcommand{\bea}{\begin{eqnarray}}
\newcommand{\beas}{\begin{eqnarray*}}
\newcommand{\eea}{\end{eqnarray}}
\newcommand{\eeas}{\end{eqnarray*}}
\newcommand{\ba}{\begin{array}}
\newcommand{\ea}{\end{array}}

\def\ls{\mathrel{\lower4pt\vbox{\lineskip=0pt\baselineskip=0pt
           \hbox{$<$}\hbox{$\sim$}}}}
\def\gs{\mathrel{\lower4pt\vbox{\lineskip=0pt\baselineskip=0pt
           \hbox{$>$}\hbox{$\sim$}}}}

\begin{document}


\title{Leptonic CP phases near the $\mu-\tau$ symmetric limit}

\author{Diana C. Rivera-Agudelo~$^1$\footnote{email: drivera@fis.cinvestav.mx},
 Abdel P\'erez-Lorenzana~$^{1,2}$\footnote{email: aplorenz@fis.cinvestav.mx}}

\affiliation{
$^1$Departamento de F\'{\i}sica, Centro de Investigaci\'on y de Estudios
Avanzados del I.P.N.\\
Apdo. Post. 14-740, 07000, M\'exico, Distrito Federal, M\'exico\\
$^2$ Facultad de Ciencias F\'{\i}sico-Matem\'aticas \\
Benem\'erita Universidad Aut\'onoma de Puebla,
72570, Puebla, Pue., M\'exico}

\begin{abstract}

The neutrino masses and mixings indicated by current neutrino oscillation 
experiments suggest that the neutrino mass matrix possesses an 
approximate $\mu-\tau$ exchange symmetry. In this study, we explore the neutrino parameter space 
and show that if a small $\mu-\tau$ symmetry breaking is considered, the 
Majorana $CP$ phases must be unequal and non-zero 
independently of the neutrino mass scale. 
Moreover, a small $\mu-\tau$ symmetry breaking favors 
quasi-degenerate masses. We also show that Majorana 
phases are strongly correlated with the Dirac $CP$ violating phase. 
Within this framework, we obtain robust predictions for the values of the Majorana phases
when the experimental indications for the Dirac $CP$ phase are used.
 
 \end{abstract}

\maketitle
\section{Introduction}

Neutrino oscillation experiments performed in the last decades have provided 
remarkable information about neutrino mixing parameters.
Global fits obtained 
with all three standard flavor neutrinos indicate that~\cite{nuglobal,PDG}, 
within 
the $3\sigma$ 
range, the mixing angles are given by solar
$\sin^2\theta_{12} \equiv \sin^2\theta_\odot\approx 0.308^{+0.064}_{-0.055}$, 
atmospheric
$\sin^2\theta_{23}\equiv \sin^2\theta_{ATM}\approx 0.437^{+0.189}_{-0.063}~ 
(0.455^{+0.186}_{-0.075})$, and reactor $\sin^2\theta_{13}\approx 
0.0234^{+0.0061}_{-0.0058} ~ (0.0240^{+0.0058}_{-0.0062})$, for the normal 
(inverted) hierarchy. The squared mass differences are also known with very 
high accuracy, where these are given by the so-called solar scale $\Delta m_{21}^2= 
m_2^2 -m_1^2\equiv\Delta 
m_{sol}^2 \approx 7.54 \times 10^{-5} eV^2$ and the atmospheric scale 
$|\Delta m_{31}^2| \sim \Delta 
m_{ATM}^2 \sim 2.4\times 
10^{-3} eV^2$. However, 
the sign in $\Delta m_{31}^2= m_3^2 -m_1^2$, and thus the neutrino mass 
hierarchy 
pattern, is still unknown. 
A very recent global analysis of neutrino oscillation data was provided by \cite{Capozzi}.

Unlike the quark sector where the mixing angles are all small, the mixings measured
in oscillation experiments are large, except for $\theta_{13}$, which was 
found to be rather small but certainly non-zero. In the standard 
parameterization, the mixings are 
given by the Pontecorvo-Maki-Nakagawa-Sakata (PMNS) matrix \cite{Pontecorvo-57-58, 
MNS},
\begin{equation}
\label{PMNSmatriz}
U_{PMNS} =  \left( \begin{array}{ccc} 
c_{12} c_{13} & s_{12}c_{13} & s_{13} e^{-i\delta_{CP}} \\
- s_{12} c_{23} - c_{12}s_{23}s_{13}e^{i\delta_{CP}}   & 
c_{12}c_{23} - s_{12}s_{23}s_{13}e^{i\delta_{CP}} & s_{23}c_{13} \\
s_{12}s_{23} - c_{12}c_{23}s_{13}e^{i\delta_{CP}} & 
- c_{12}s_{23} -c_{23}s_{12}s_{13}e^{i\delta_{CP}} & c_{23}c_{13} 
\end{array} \right)\cdot K ~, 
\end{equation}
where $c_{ij}$ and $s_{ij}$ denote $\cos \theta_{ij}$ and 
$\sin \theta_{ij}$, respectively, and the mixing angles are given as 
$\theta_{12}$, $\theta_{13}$, and $\theta_{23}$, where $\delta_{CP}$ is the
Dirac \textit{CP}-violating phase, whereas
$K=\text{Diag}(e^{-i\beta_1/2},e^{-i\beta_2/2}, 1)$ is a diagonal matrix 
containing two Majorana \textit{CP}-violating phases, 
$\beta_1$ and $\beta_2$, which do not contribute to neutrino oscillations. 
In addition, although $\delta_{CP}$ has not been 
determined well, several suggestions from global fits 
~\cite{nuglobal,PDG,Capozzi,fitvalle} support 
$\delta_{CP}\sim -\pi/2$. 
A number of ongoing and 
future oscillation neutrino experiments aim to determine the 
neutrino mass hierarchy and the Dirac \textit{CP}-violating phase. 
Determining these parameters will be very important for identifying the 
flavor symmetry 
that underlies the pattern of lepton flavor mixing. 

After considering a basis where the charged lepton masses and weak interactions are simultaneously 
diagonal, the labels associated with weak flavors, $\alpha, \beta = e,\mu,\tau$, 
are transparent and the PMNS
matrix also becomes that which diagonalizes the neutrino mass terms, 
where they are
generally given by the effective operator
\be
 \bar{\nu}_{\alpha L}(M_\nu)_{\alpha\beta} \nu_{\beta L}^c 
+ h.c., 
\ee
such that $U_{PMNS} = U_\nu\cdot K$. Therefore, the neutrino mass matrix can be
written in terms of a diagonal (complex) mass matrix, 
$M_{diag} = \text{Diag}\{ m_1 e^{-i\beta_1},m_2 e^{-i\beta_2}, m_3 \}$,  
simply as
\be 
M_\nu = U^\ast_\nu\cdot M_{diag} \cdot U_\nu^\dagger.
\label{mnu}
\ee
In the following, we employ this formulation and we 
denote
$m_{j}\equiv |m_{j}| e^{-i\beta_j}$, for $j=1,2$. 

It should be noted that while the observed 
$\theta_{13}$ is small but not zero, $\theta_{ATM}$ 
is close to its maximal value, $\pi/4$. 
Clearly, neither of the central values of these angles are the exact critical
values ($\theta_{13}=0$ and $\theta_{ATM}=\pi/4$), but it is intriguing to observe that it is possible to establish an approximate empirical relation
regardless of the hierarchy, 
\be
1/2 - \sin^2\theta_{ATM}\approx \sin\theta_{13}/\text{few}~, 
\label{phen}
\ee
which suggests that the deviation of $\theta_{ATM}$ 
from its maximal value, $\Delta\theta=\pi/4 -|\theta_{ATM}|$, could 
be correlated to the deviation from zero present in 
$\theta_{13}$~\cite{APL-DRA}.
In these terms, we note that the empirical relation given above can 
be simply rewritten as $\sin\Delta\theta\approx \sin\theta_{13}/\text{few}$.
Theoretically, this may facilitate an understanding 
of the observed mixings as possibly having a common physical 
origin by highlighting a well-defined flavor symmetry. 
In fact, it is easy to see that null values of $\Delta\theta$ and 
$\theta_{13}$ do increase the symmetry of the neutrino sector.  
These values imply that the mixing matrix $U_\nu$ will take the bimaximal 
form
\begin{equation}\label{Ubm}
U_{BM}=  \left( \begin{array}{ccc} 
\cos\varphi_{12}                     & \sin\varphi_{12}                  & 0 \\
\frac{- \sin\varphi_{12}}{\sqrt{2}}  & \frac{\cos\varphi_{12}}{\sqrt{2}} & 
\frac{-1}{\sqrt{2}} \\
\frac{- \sin\varphi_{12}}{\sqrt{2}}  & \frac{\cos\varphi_{12}}{\sqrt{2}} &  
\frac{1}{\sqrt{2}}
\end{array}\right) ,
\end{equation}
where the only undefined angle is $\varphi_{12}$, 
which could be taken as the solar mixing. If Eq.~(\ref{Ubm}) is combined with 
Eq.~(\ref{mnu}) and 
the mass matrix elements are defined as $m^0_{\alpha\beta} = (M_\nu)_{\alpha\beta}$, then we 
obtain
\bea
m^0_{ee}&=& m_{1}\cos^{2}\varphi_{12}+m_{2}\sin^{2}\varphi_{12},\nonumber\\
m^0_{e\mu}=m^0_{e\tau}&=&\frac{\sin2\varphi_{12}}{\sqrt{8}}\left(m_{2}-m_{1}
\right),\nonumber\\
m^0_{\mu\tau}&=&\frac{1}{2}\left(m_{1}\sin^2\varphi_{12}+m_{2}\cos^2\varphi_{
12} -m_ { 3 } \right),\nonumber\\
m^0_{\mu\mu}=m^0_{\tau\tau}&=&
\frac{1}{2}\left(m_{1}\sin^{2}\varphi_{12}+m_{2}\cos^2 \varphi_{12} 
+m_{3}\right)~.\label{mterms}
\eea
The two general conditions given as 
$m^0_{e\mu}=m^0_{e\tau}$ and $m^0_{\mu\mu}=m^0_{\tau\tau}$ reduce the number of 
free mass parameters to four and give rise to the so-called $\mu - \tau$ 
symmetry~\cite{mutau}. As a consequence, the observed values of $\theta_{13}$ 
and $\Delta \theta$ can be understood as a result of the breaking of 
$\mu-\tau$ symmetry.
This issue has inspired many theoretical studies in recent 
years~\cite{APL-DRA,mutau,others,mtseesaw}. A very recent review of the status 
of the $\mu-\tau$ flavor symmetry was provided by \cite{revmutau}. 

However, there have been very few studies of the possible values for CP 
violating phases (\textit{CPVP}) in this context.
In particular, \cite{predictingcp} provides different values for the Dirac phase $\delta_{CP}$ considering small deviations from the $\mu-\tau$ symmetry, although the Majorana phases were not considered.
 To the best of our 
knowledge, the conditions under which $\mu-\tau$ 
symmetry breaking is small have not been explored. 
This could be relevant because small symmetry 
breaking is considered as the starting point in this type of study. In 
this study, we provide a detailed analysis of the magnitude of the $\mu-\tau$ 
symmetry breaking using the current data for neutrino mixing 
parameters, and we identify the mass spectrum and \textit{CPVP} required to obtain 
small symmetry breaking. 

The remainder of this paper is organized as follows. First, we parameterize the breaking of 
$\mu-\tau$ symmetry. Next, we use experimental results obtained for neutrino masses and 
mixings to explore the breaking parameter space but without assuming any special values
for \textit{CPVP}. We then show that relatively small 
parameters, and thus 
perturbative approximations, are still allowed by the data only in specific 
cases. In particular, we show that $\mu-\tau$ symmetry breaking can be small 
(less than 10\%) when the neutrino masses are 
almost degenerate and for specific 
values of \textit{CPVP} that are strongly correlated. We also show that 
this correlation indicates a well-defined region in the parameter space for 
\textit{CPVP}, which is a testable feature for models near the $\mu-\tau$ 
symmetric limit.
Finally, we present 
our conclusions.

\section{$\mu -\tau$ symmetry breaking parameters}\label{secbreaking}
As mentioned above, experimental results show that 
$\theta_{13}$ and $\Delta\theta$ are non-zero, and 
thus the 
$\mu-\tau$ symmetry is broken. However, this breaking can actually be small and in some cases,
the $\mu-\tau$ symmetry may be considered as an 
approximate symmetry. In fact, \cite{gupta} found that very 
small symmetry breaking can be possible in the quasi-degenerate hierarchy. 
In this study, we consider the breaking of $\mu-\tau$ 
to identify 
the hierarchy, as well as extracting some useful information about \textit{CPVP}, 
which supports small symmetry breaking. In general, any generic neutrino 
mass matrix can always be parameterized in terms of a symmetric part plus a 
correction that explicitly breaks the symmetry by 
$M_\nu = M_{\mu - \tau} + \delta M,$ where $ M_{\mu-\tau}$ has a 
$\mu-\tau$ symmetry and $\delta M$ is defined by only two non-zero 
elements, 
\begin{equation}
\delta M = \left( \begin{array}{ccc} 
0  & 0  & \delta  \\
0 & 0 & 0 \\
\delta & 0 & \epsilon
\end{array} \right)~, 
\end{equation}
where the breaking parameters are clearly defined as $\delta = m_{e\tau} - 
m_{e\mu}$ and $\epsilon = m_{\tau \tau} - m_{\mu \mu}$. 
Of course, there is no a priori reason why these parameters should be small 
compared with $M_{\mu-\tau}$ mass elements.
In order to deal with dimensionless parameters, we define
\begin{eqnarray}
\label{deltaepsilon2}
 \hat \delta \equiv \frac{\delta}{m_{e \mu}} &= 
\frac{ \sum_i (U^*_{ei} U^*_{\tau i}-  U^*_{ei} U^*_{\mu i} ) m_i}{\sum_i U^*_{ei} U^*_{\mu 
i} m_i },
 \nonumber \\
\hat \epsilon \equiv \frac{\epsilon}{m_{\mu \mu}} &= 
\frac{\sum_i (U^*_{\tau i} U^*_{\tau i}-  
U^*_{\mu i} U^*_{\mu i}) m_i}{\sum_i U^*_{\mu i} U^*_{\mu i} m_i}~,
\end{eqnarray}
where the right-hand-sides are written according to Eq.~(\ref{mnu}). As 
usual, we take $i = 1,2,3$.
The latter expressions give the complex parameters,
$\hat \delta$ and $\hat \epsilon$, in terms of three observed mixing angles, 
three absolute masses, $|m_{1,2,3}|$, and three \textit{CPVP}.

For the purposes of calculation, we rewrite the absolute masses in 
terms of the two observed mass squared differences involved in neutrino 
oscillations, $\Delta m^2_{sol}$ and $\Delta m^2_{ATM}$, and the lightest 
absolute neutrino mass, i.e., $m_0$, as
\begin{eqnarray}
\label{eigenvalues}
|m_2| = \sqrt{m_0^2 +  \Delta m_{sol}^2} ~&\text{~and}& ~
|m_3|= \sqrt{m_0^2 + |\Delta m_{ATM}^2|}~~ \text{for NH}. \\
|m_1| = \sqrt{m_0^2 + |\Delta m_{ATM}^2|} ~&\text{~and}& ~
|m_2|= \sqrt{m_0^2 +| \Delta m_{ATM}^2|+ \Delta m_{sol}^2} ~~
\text{ for IH}. \nonumber
\end{eqnarray}
Note that $m_0$ becomes 
$|m_1|$ for the normal mass hierarchy (NH) and $m_3$ 
for the inverted 
mass hierarchy (IH). 
Thus, the breaking parameters, $\hat \delta$ and $\hat \epsilon$, 
depend 
on nine observables: three mixing angles, two mass squared 
differences, the lightest neutrino mass, and three \textit{CPVP}, where 
neutrino oscillation experiments have already provided accurate values for 
the first five, so only the last four are 
unknown. Hence, in 
practical terms, the breaking parameter space $\hat\delta-\hat\epsilon$ that 
can accommodate neutrino mixings and oscillation mass scales within 
experimental uncertainties should in principle remain 
undetermined due to the arbitrariness of the $m_0$ and $CP$ phases. 
However, we suggest that by assuming the smallness of $|\hat\delta|$ 
and $|\hat \epsilon|$, the resulting bounded parameter space 
will not be
consistent with any arbitrary values of the neutrino 
observables, but instead it will yield
specific predictions for the lightest neutrino mass and $CP$ phases. We 
elaborate on this idea in the following.

We perform a general 
scan of the absolute values of $\hat \delta$ and $\hat \epsilon$ 
using the data obtained from oscillation experiments, where we allow $m_0$ and 
\textit{CPVP} to vary within $(0-0.4)$~eV and $(0-2\pi)$, respectively. 
However, before presenting our numerical results, we perform an approximate
analysis of the expressions in Eq.~(\ref{deltaepsilon2}) in order to obtain 
some insights into the expected conditions required for small symmetry 
breaking. These expressions can be written in a suitable form as 
\begin{eqnarray}\label{dehat}
\hat{\delta} &=& \frac{y_{-} f  s_{13}-y_{+}}{1+f \ s_{13} \tan~\theta_{23}}~,  
\nonumber \\
\hat{\epsilon} &=& \frac{g \cos~2\theta_{23} - s_{13}h}{1+g s_{23}^2+s_{13}h/2} ~,
\end{eqnarray}
where
\begin{eqnarray}
y_{\pm} &=& \frac{c_{23}\pm s_{23}}{c_{23}}, \nonumber \\
f &=& \frac{(c_{12}^2 |m_1| e^{-i\beta_1} + s_{12}^2 |m_2| 
e^{-i\beta_2})e^{-i\delta_{CP}} -m_3 e^{i\delta_{CP}} }
  {c_{12}s_{12}(|m_1| e^{-i\beta_1} - |m_2| e^{-i\beta_2} )}~, 
\nonumber \\
g &=&\frac{( c_{12}^2 s_{13}^2 - s_{12}^2) |m_1| e^{-i\beta_1}+( s_{12}^2 
s_{13}^2 -c_{12}^2) |m_2| e^{-i\beta_2} + m_3 c_{13}^2 }
{s_{12}^2  |m_1| e^{-i\beta_1} +c_{12}^2  |m_2| e^{-i\beta_2}}~, 
\nonumber\\
h &=&\frac{(|m_1| e^{-i\beta_1} - |m_2|  e^{-i\beta_2} )\sin~2\theta_{23}~ 
\sin~2\theta_{12}e^{-i\delta_{CP}}}
{s_{12}^2 |m_1| e^{-i\beta_1} + c_{12}^2 |m_2| e^{-i\beta_2}} ~.
\label{fgh}
\end{eqnarray}
As shown by these expressions, $\hat \delta$ and $\hat \epsilon$ in 
Eq.~(\ref{dehat}) become zero when $\theta_{13}=0$ and $\theta_{23}=-\pi/4$, as 
expected from the exact $\mu-\tau$ symmetry. 
Next, let us analyze the conditions for $|\hat \delta|$,$|\hat 
\epsilon| \ll 1$. 
According to the three possible approaches given by NH and IH,
the almost degenerate limit where all neutrino masses are 
about the same order,
and using the 
central values for the 
current mixing parameters~\cite{nuglobal}, we have the following.
\begin{itemize}
\item  For NH, $|m_1| \ll |m_2| \approx \sqrt{\Delta m_{sol}^2} \ll m_3 
\approx \sqrt{\Delta m_{ATM}^2}$, and thus
\begin{equation}
f \approx \frac{e^{i(\delta_{CP}+\beta_2)}}{ s_{12}~c_{12}} 
\sqrt{\frac{\Delta m_{ATM}^2}{\Delta m_{sol}^2}} 
\left(1-\mathcal{O} \left( \frac{\Delta m_{sol}^2}{\Delta m_{ATM}^2}\right) 
\right)~, 
~|f|\sim 12.5 ~,
\end{equation} 
which implies that $|\hat \delta| \sim 3.26$. Thus, in NH, the breaking 
of $\mu-\tau$ symmetry by $\hat \delta$ is always large, and thus
there is no need to examine $\hat \epsilon$. We note that this conclusion is 
independent of the values of \textit{CPVP}. These results are confirmed by 
the numerical analysis presented below.  

\item For IH, we have 
$|m_1|\approx \sqrt{\Delta m_{ATM}^2}~,~  
|m_2| \approx \sqrt{\Delta m_{sol}^2+ \Delta m_{ATM}^2} \gg m_3 $, 
which gives
\begin{equation}
f \approx \frac{-e^{-i\delta_{CP}}(c_{12}^2 e^{-i(\beta_1-\beta_2)} + s_{12}^2 + 
\frac{s_{12}^2}{2}\frac{\Delta m_{sol}^2}{\Delta 
m_{ATM}^2})}{s_{12}~c_{12}\left(1- e^{-i(\beta_1-\beta_2)}+ \frac{1}{2} \frac{\Delta 
m^2_{sol}}{\Delta m^2_{ATM}} \right)} ~. 
\end{equation}
In this case, we can see that $|f|$ is very large when $\beta_1-\beta_2=0$, 
whereas $|f|\sim 1$ when $\beta_1-\beta_2=\pm \pi$, and thus 
$|\hat \delta|\sim 0.1$, which is desirable. In addition, 
$\hat \epsilon$ is given in terms of $g$ and $h$, as in Eq. (\ref{dehat}), which 
 can now be approximated as \begin{equation}
g\approx\frac{( c_{12}^2 s_{13}^2 - s_{12}^2)e^{-i(\beta_1-\beta_2)}+
( s_{12}^2 s_{13}^2 
-c_{12}^2)}{s_{12}^2e^{-i(\beta_1-\beta_2)} +c_{12}^2} ~,
\end{equation}
\begin{equation}
h\approx\frac{(e^{-i(\beta_1-\beta_2)} -1)\sin~2\theta_{23}~ 
\sin~2\theta_{12}e^{-i\delta_{CP}}}{s_{12}^2 e^{-i(\beta_1-\beta_2)} +c_{12}^2} ~.
\end{equation}
From the expressions above, we can see that $\beta_1-\beta_2=\pm \pi$ gives 
$g\approx -1$ 
and $h\approx 4e^{-i\delta_{CP}}$, from which we 
obtain $|\hat \epsilon|\sim 0.6$. Hence, the largest 
contribution to $\mu-\tau$ breaking in this case comes from $|\hat \epsilon|$ 
rather than $|\hat \delta|$.

\item Finally, in the degenerate hierarchy (DH) limit, where $|m_1|\approx |m_2| \approx m_3$, we obtain\\
\begin{equation}\label{fdh}
f\approx 
\frac{-e^{-i\delta_{CP}}(c_{12}^2 e^{-i(\beta_1-\beta_2)} + 
s_{12}^2)-e^{i\delta_{CP}}e^{i\beta_1}}
{s_{12}~c_{12}(1- e^{-i(\beta_1-\beta_2)})} ~.
\end{equation}
\begin{equation}\label{gdh}
g \approx \frac{( c_{12}^2 s_{13}^2 - s_{12}^2)e^{-i(\beta_1-\beta_2)}
+( s_{12}^2 s_{13}^2 -c_{12}^2) + c_{13}^2 e^{-i\beta_1 }}
{s_{12}^2 e^{-i(\beta_1-\beta_2)} +c_{12}^2}~, 
\end{equation}
\begin{equation}\label{hdh}
h \approx \frac{(e^{-i(\beta_1-\beta_2)} - 1 )\sin~2\theta_{23}~ 
\sin~2\theta_{12}e^{-i\delta_{CP}}}{s_{12}^2 e^{-i(\beta_1-\beta_2)} +c_{12}^2} ~.
\end{equation}
From these equations, we again note that $|f|$ is strongly enhanced for 
$\beta_1-\beta_2=0$, which implies that $|\hat \delta|\textgreater 1$, e.g., 
for $\beta_1=\beta_2=\pi$ and $\delta_{CP}=-\pi/2$, we obtain $|f|\sim 68$, which 
leads to $|\hat \delta| \sim 2$. However, other specific 
values of \textit{CPVP} may lead to small breaking parameters, e.g., 
for 
$\beta_1=\pi$, $\beta_2=\pi/2$ and $\delta_{CP}=-\pi/2$, we obtain $|f|\sim 0.6$, 
$|g|\sim 2$, and $|h|\sim 1.7$, such that $|\hat \delta| \sim 0.2$ and $|\hat 
\epsilon| \sim 0.1$.     

\end{itemize}

Based on the previous analysis, we can conclude that the case of equal Majorana 
phases is strongly disfavored in any hierarchy. In addition, we can easily determine 
that the symmetry is strongly broken for the NH. This is consistent with the 
previous results presented by~\cite{APL-DRA}, where only \textit{CP} 
conserving situations were analyzed. It is also consistent with the numerical 
analysis presented in the following.

\begin{figure}[t!]
\begin{center}
\includegraphics[scale=0.8]{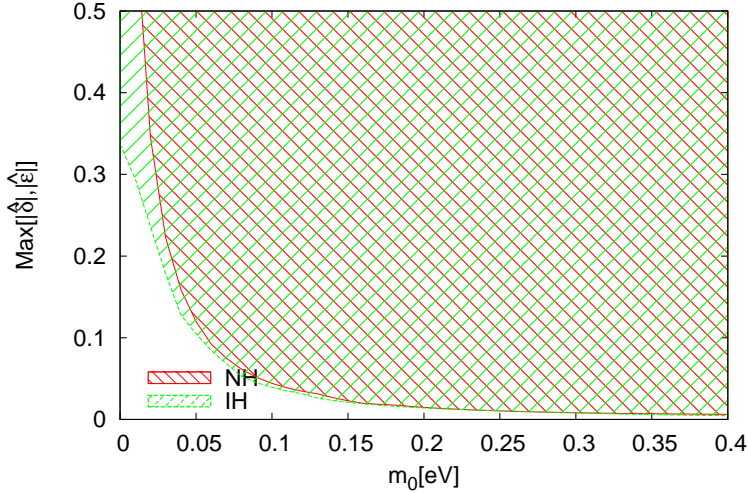}
\end{center}
\caption{\label{brakingspace} 
\small{Allowed region (hashed area) based on the experimental 
data for Max[ $|\hat \delta|$, $|\hat 
\epsilon|$], and the maximum between $|\hat \delta|$ and $|\hat \epsilon|$ as a 
function of the lightest neutrino mass, $m_0$, for NH and 
IH.
The regions are obtained by varying $\delta_{CP}$ and 
$\beta_{1,2}$ within the interval $[0,2\pi]$.
} }
\end{figure} 

In Fig.~\ref{brakingspace}, we show the allowed region for the 
symmetry breaking parameters obtained by varying the Dirac phase, 
$\delta_{CP}$, and Majorana phases, $\beta_1$ and $\beta_2$, 
within the $(0-2\pi)$ interval, but taking the mixing parameters within a
3 $\sigma$ level. We depict the
maximum value between $|\hat \delta|$ and $|\hat \epsilon|$, i.e., Max[$| \hat 
\delta|$,$| \hat \epsilon|$], as a function of the lightest neutrino mass. According to 
this figure, we can easily see that for $m_0\sim 0$~eV in the NH, the 
breaking parameters 
are very large regardless of the \textit{CPVP} values, as found in 
the previous 
analysis. However, for $m_0\sim 0$~eV in the IH, the 
breaking 
is at least of 30$\%$. 
Finally, and more interestingly, we can see that for 
$m_0\gtrsim 0.05$~eV, 
which 
corresponds to the 
almost DH, the breaking can be less 
than 10$\%$. This indicates that DH is 
the best hierarchy for regarding $\mu-\tau$ as a good approximate 
symmetry, which is consistent with the results obtained by \cite{gupta} using a different approach.

However, it should be noted that even larger 
values for the breaking parameters are possible in the DH case, which is 
clear from the figure. Imposing the phenomenological requirement of a very
small amount of breaking, 
as suggested by the theoretical indication of a perturbative origin, 
has the effect of cutting off all the parameter space above 
$\text{Max}[|\hat\delta|,| \hat \epsilon|]>0.1$, or any other selected small 
value. Of course, this can be achieved by excluding the values for 
\textit{CPVP} that are consistent only with larger breaking 
of the symmetry. Therefore, we may conclude that a small $\mu-\tau$
symmetry breaking implies more specific values for the $CP$ phases, as 
discussed in the following.

\section{Majorana phases and $\mu -\tau$ symmetry}

In the previous section, we showed that the current experimental data indicate 
that the requirement for small symmetry breaking favors the 
quasi-degenerate mass hierarchy, i.e., $m_0\gtrsim 0.05$~eV.
Accordingly, we take $m_0=0.1$~eV 
in the following. However, we note that our results do not change significantly if 
we consider other values for $m_0$ within that range. This can be understood 
as a consequence of the fact that Eqs.~(\ref{fdh}) to 
(\ref{hdh}), which ultimately define 
the breaking parameters (see Eq.~\ref{dehat}), have no explicit dependence on 
the absolute neutrino mass scale in this limit. Nevertheless, 
we use the exact expressions in (\ref{dehat}) and (\ref{fgh}) for 
our numerical analysis, thereby avoiding any further approximation.

\begin{figure}[t!]
\begin{center}
\includegraphics[scale=0.6]{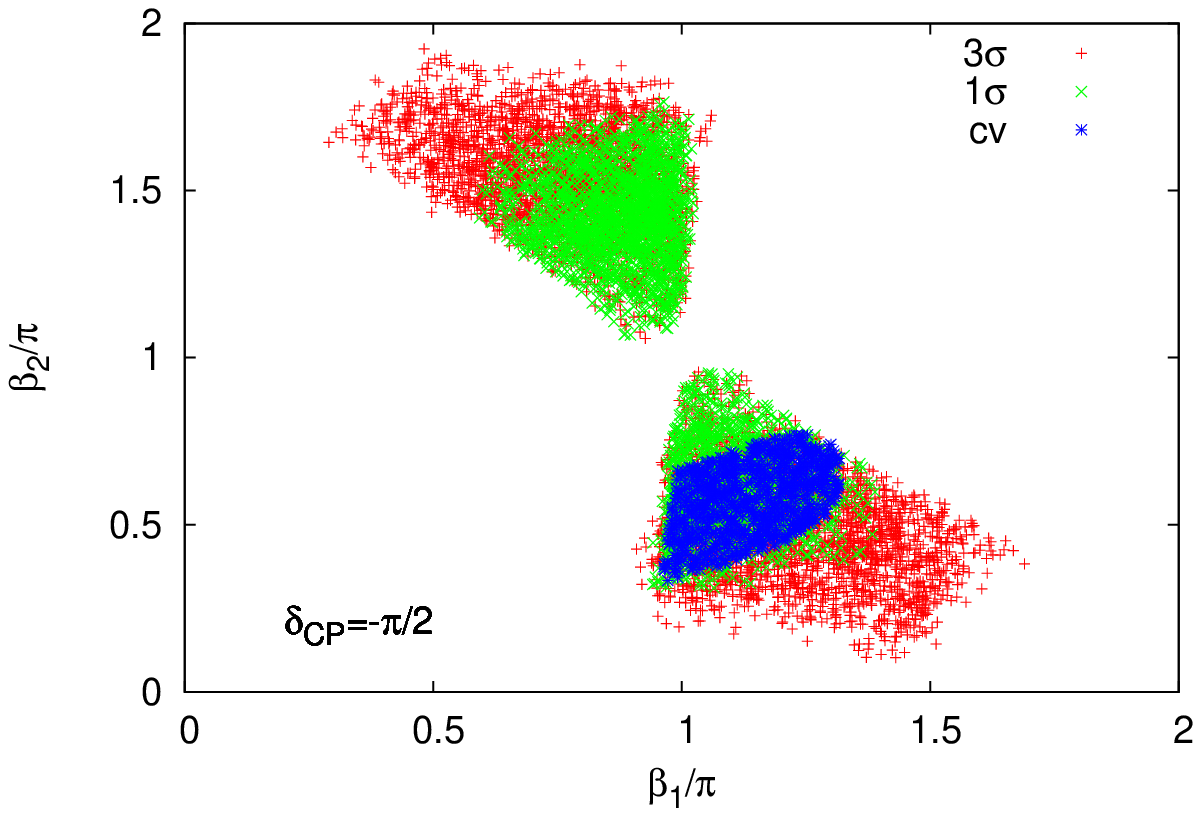}~
\includegraphics[scale=0.6]{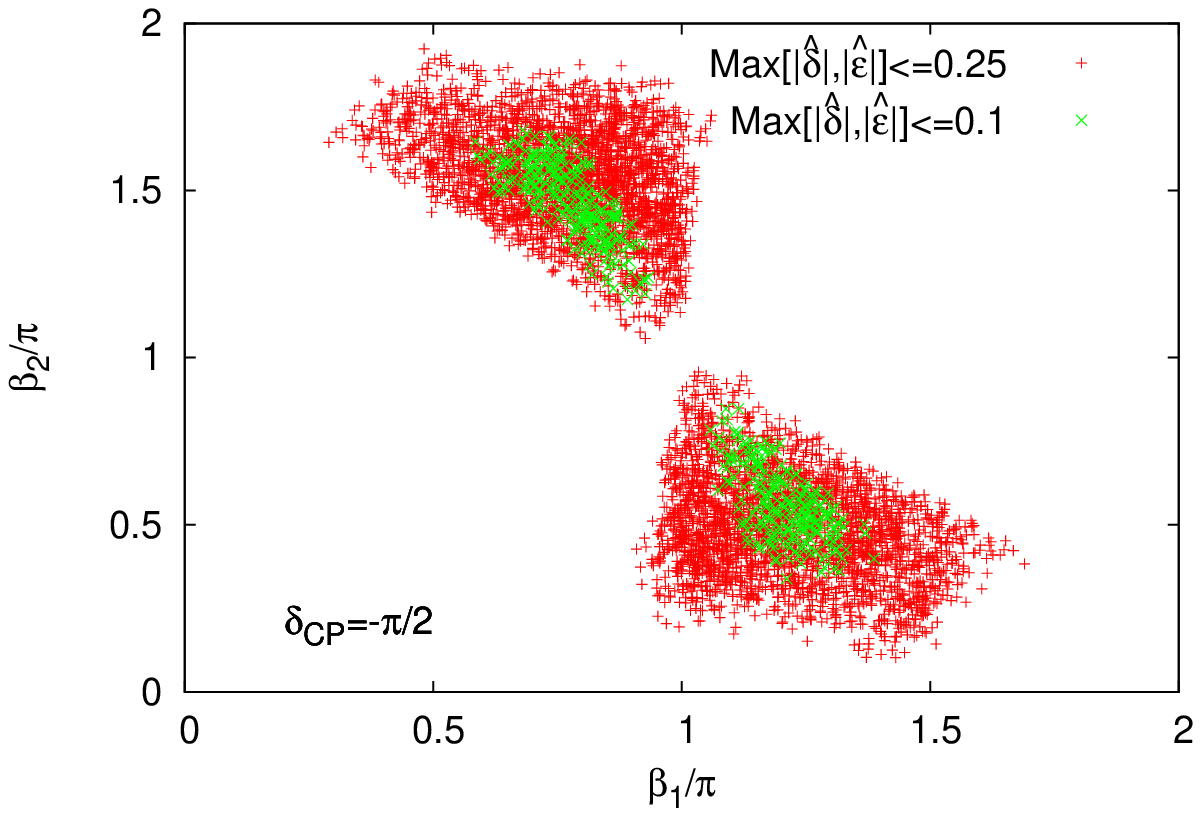}
\end{center}
\caption{\label{maj1} 
\small{Allowed regions for the two Majorana phases for $\delta_{CP}=-\pi/2$. 
In the left-hand plot, the green color (cross signs) corresponds to the mixing 
parameters 
varied up to 1$~\sigma$ error, the red color (plus signs) corresponds to the 
3$~\sigma$ interval, and the blue color (stars) corresponds to the central values 
(cv) of the mixing parameters, which match with 
Max[$|\hat \delta|$, $|\hat\epsilon|$]$\leq 0.25$. In the right-hand plot, the 
green color (cross signs) 
corresponds to Max[$|\hat \delta|$, $|\hat \epsilon|$]$\leq 0.1$ and the red 
color 
(plus signs) corresponds to Max[$|\hat \delta|$, $|\hat \epsilon|$]$\leq 0.25$, where the 
experimental values are varied within 3~$\sigma$. 
} }
\end{figure} 

After selecting $m_0$, 
$\hat \delta$ and $\hat \epsilon$ only depend on 3 \textit{CPVP}. 
Thus, by selecting a suitable value of the Dirac phase, $\delta_{CP}$, we 
can obtain some information about the two remaining Majorana phases that 
fulfill the requirement for small symmetry breaking. In summary, any two 
given values for the phases will provide a unique set of absolute breaking 
parameters, and vice versa. Therefore, only a well-defined 
allowed region in the CP phase space will be consistent with our symmetry breaking 
requirement.
The results of our numerical analysis show this and they are depicted 
in Fig.~\ref{maj1}, where the left-hand side shows the allowed 
region for the two Majorana phases, $\beta_1$ and $\beta_2$, which match with 
the condition that Max[$|\hat \delta|$, $|\hat \epsilon|$]$\leq 0.25$. To 
explain this plot, we select $\delta_{CP}=-\pi/2$, which is near to 
the best fit 
value reported by~\cite{nuglobal,PDG,Capozzi,fitvalle}. The other mixing parameters are 
varied up to 1 $\sigma$ and 3 $\sigma$ from their 
central values, respectively. The corresponding regions 
are indicated in the figure.
We can easily see that there is a strong correlation between the two 
Majorana phases, which can be inferred from the expressions in Eqs. 
(\ref{fdh} --\ref{hdh}), where the relative phase $\beta_1-\beta_2$ appears. 
This figure shows that the values of
$\beta_1=\pi$ and 
$\beta_2=\pi/2$ discussed in the previous section, which lead to 
Max[$|\hat \delta|$, $|\hat \epsilon|$]$\approx 0.2$, are contained well within 
the allowed regions. An interesting outcome that we want 
to stress is that null
Majorana phases are excluded completely at the 3$\sigma$ level.
Furthermore, 
as shown by the same plot, the allowed region 
is not modified greatly when we move the experimental values from 
1$\sigma$ to 
3$\sigma$ deviation limits. Thus, the correlation is only slightly 
sensitive 
to small mixing angle variations. By contrast, as shown 
in the right-hand 
plot in Fig. \ref{maj1}, when we restrict 
the upper bound condition even more on 
Max[$|\hat \delta|,|\hat \epsilon|$], then 
the allowed region for $\beta_1$ and 
$\beta_2$ is reduced greatly. Therefore, in this framework, it is 
possible to obtain some indications about the Majorana phases, as well as the pattern of masses, according to the analysis in 
the last section. 
We consider that this would be useful for future research 
based on an approximate $\mu-\tau$ symmetry. 

\begin{figure}[t!]
\begin{center}
\includegraphics[scale=0.6]{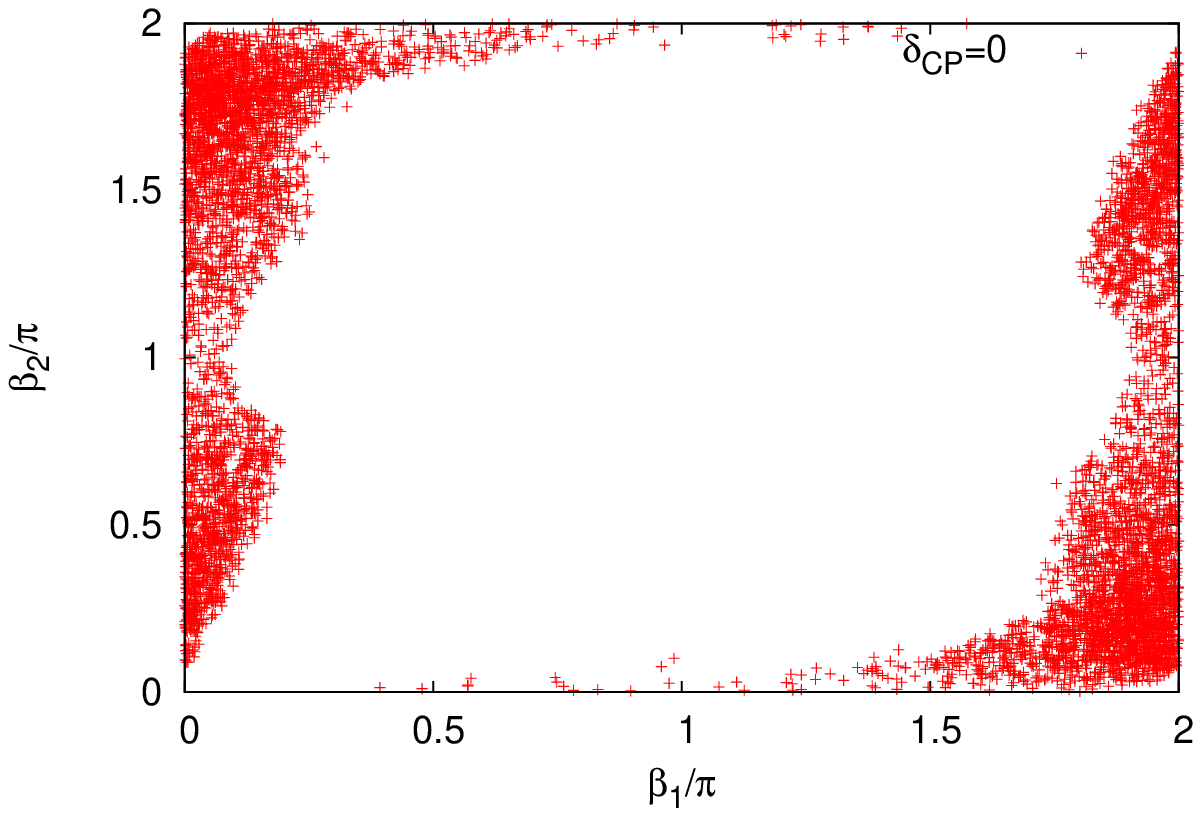}~
\includegraphics[scale=0.6]{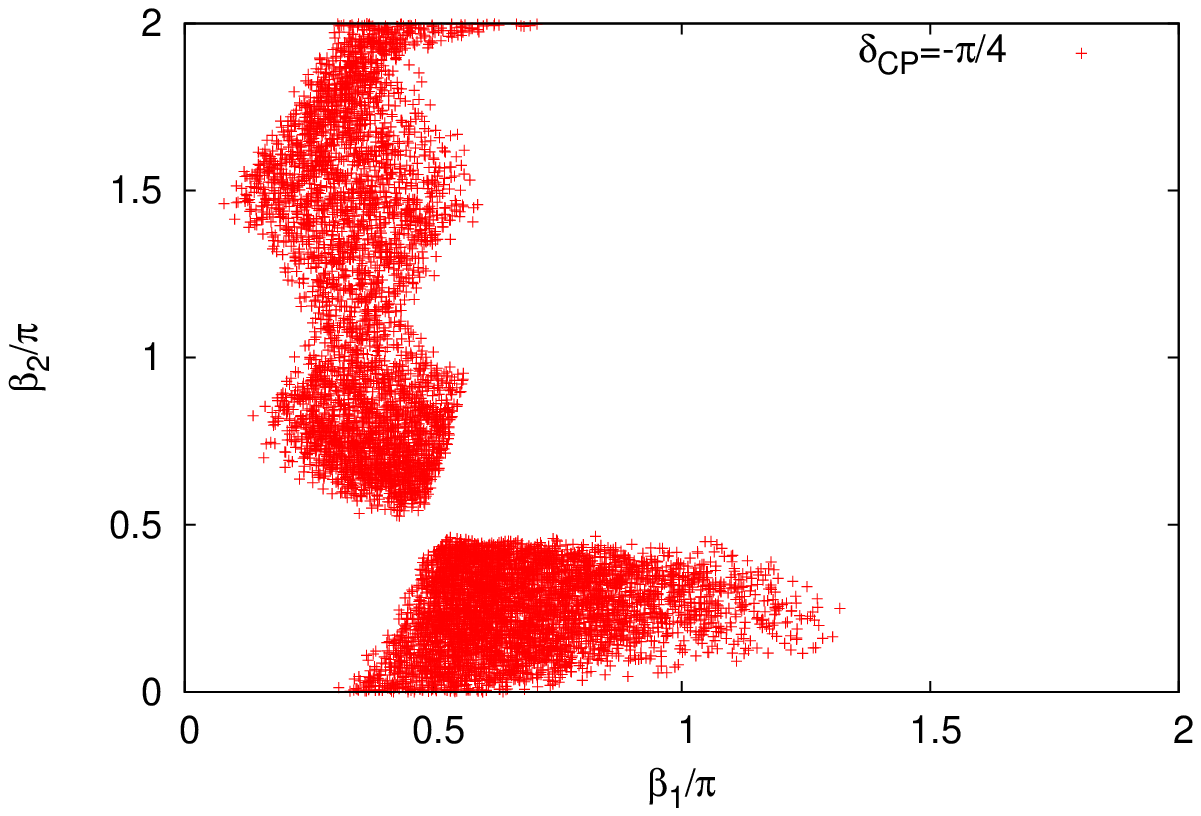}
\end{center}
\caption{\label{maj2} 
\small{2$\sigma$ allowed regions for the Majorana phases, $\beta_1$ and 
$\beta_2$, for 
Max[$|\hat \delta|$, $|\hat \epsilon|$]$\leq 0.25$. The left-hand plot 
corresponds to $\delta_{CP}=0$ and the right-hand plot to $\delta_{CP}=-\pi/4$. 
} }
\end{figure}  

Finally, in order to study the dependence of our results on the 
value selected for 
the Dirac phase, we consider different values in our analysis, thereby demonstraing that there is 
a strong correlation between the allowed Majorana phase space and the Dirac $CP$ 
phase.
Two distinctive output examples are presented in Fig.~\ref{maj2}, 
where the
values of $\beta_1$ and $\beta_2$ are shown that match the conditions
Max[$|\hat \delta|$, 
$|\hat \epsilon|$]$\leq 0.25$, for $\delta_{CP}=0$ and $\delta_{CP}=-\pi/4$ 
(left- and right-hand sides, respectively).
From Figs.~\ref{maj1} and \ref{maj2}, we can seen that in any of these 
cases, equal Majorana phases are totally disfavored for small breaking, as 
discussed in Sec.~\ref{secbreaking}. In particular, simultaneously 
zero
Majorana phases remain excluded in any of the cases considered.
In addition, 
we note that the two Majorana phases are 
sensitive to variations in the Dirac 
\textit{CP} phase.   
This is expected from Eqs.~(\ref{fdh}-\ref{hdh}), where
$\delta_{CP}$ enters as a global phase factor that cannot cancel out. 
Nonetheless, the 
regions associated with different values of the Dirac $CP$ phase are 
completely different, which supports 
an interesting conclusion within the small $\mu-\tau$ breaking hypothesis, so in the 
near future, determining the Dirac $CP$ phase will provide specific 
predictions for the allowed values of the Majorana phases,
thereby making this hypothesis testable.


\section{Concluding remarks}

In this study, we provided a complete analysis of the two dimensionless parameters 
$\hat \delta$ and $\hat \epsilon$, which encode $\mu-\tau$ symmetry breaking. 
By taking neutrino oscillation mixings and mass scales according to the 
current experimental data, we showed that the breaking
parameters depend only on four free variables: the lightest neutrino 
mass, 
$m_0$, the Dirac \textit{CP} phase, $\delta_{CP}$, and the two Majorana 
phases. First, we studied these parameters in an
analytical manner under 
the three approximations given by the hierarchies. 
We found that equal 
Majorana phases lead to strong breaking of $\mu-\tau$ symmetry regardless of 
the hierarchy. We also verified the result reported by \cite{gupta}, 
thereby indicating that small 
symmetry breaking does prefer 
a quasi-degenerate mass hierarchy. 
Next, by allowing all three 
\textit{CPVP} to vary within the
[$0-2\pi$] interval, we numerically studied 
the absolute values of both the $\hat 
\delta$, and $\hat \epsilon$ parameters in order to quantify the breaking of 
$\mu-\tau$ as a function of $m_0$. Based on this analysis, we found that for 
$m_0\lesssim 0.02$~eV in the NH, the $\mu-\tau$ symmetry will be broken by 
more than a $50$\%. 
By contrast, in the case of IH, we found that for $m_0=0$~eV, the 
breaking is always larger than a 30\%. However, this limit decreases when 
$m_0$ increases regardless of the hierarchy. Clearly, this indicates that 
in only the almost degenerate neutrino masses limit, when $m_0\gtrsim 0.05$~eV, 
we can obtain breaking as small as 10\%. Therefore, we 
may conclude that 
although current neutrino data are consistent with an approximate 
$\mu-\tau$ symmetry, the very near to symmetry limit 
prefers the quasi-degenerate active neutrino spectrum.

Based on the above, we restricted the analysis to the large 
absolute neutrino mass range, i.e., for $m_0\gtrsim 0.1$~eV, and we then focused 
only on the parameter space where $|\hat\delta|$, $|\hat \epsilon|$ are 
effectively small to explore the allowed $CP$ phases. Remarkably, our analysis 
showed that the smallness of the breaking parameters is fairly independent of the 
specific value taken by $m_0$, but their size is strongly governed by the 
\textit{CPVP}. The latter appear to be strongly 
correlated among 
themselves up to the point that a given value of the 
Dirac phase should predict a well-bounded allowed parameter space for the Majorana 
phases.
The allowed values for the latter are slightly sensitive to 
the 
variations in oscillation mixings and mass scales up to 3$\sigma$ level, but as expected,
they are highly dependent on the allowed maximum values for 
the breaking parameters. 

Finally, we can make two important conclusions based on our analysis. 
First, if we regard $\mu-\tau$ 
as a good approximate symmetry, then the Majorana phases will definitely be 
non-zero. Second, a future determination of the Dirac phase will greatly 
restrict 
the Majorana phases that may be compatible with the condition of a weak 
breaking of $\mu-\tau$. It is interesting that a recent global analysis provided by 
\cite{Capozzi} confirmed the previous intriguing preference for negative values 
of $\sin(\delta_{CP})$ with a best fit for $\delta_{CP}$ near $-\pi/2$, which 
makes the robust prediction of the Majorana phases obtained for 
this value more substantial. Experimental searches
of the Dirac phase, neutrinoless double beta decay, and 
determinations of the mixing angles with improved precision would 
provide useful information to support or exclude the scenario for 
small $\mu-\tau$ symmetry breaking.

\section*{Acknowledgments}

A.P.L. thanks FCFM-BUAP for their warm hospitality. 
This study was partially supported
by CONACyT, M\'exico, under grant No.237004.



\begin{thebibliography}{99}

\bibitem{nuglobal}
F. Capozzi et al., Phys. Rev. D89, 093018 (2014).

\bibitem{PDG}
For a detailed discussion see
K.A. Olive et al. (Particle Data Group Collaboration), Chin. Phys. C, {\bf 38}, 090001 
(2014). 

\bibitem{Capozzi}
F. Capozzi et al., arXiv:1601.07777 [hep-ph].

\bibitem{Pontecorvo-57-58}
B. Pontecorvo, J. Exptl. Theoret. Phys. {\bf 33}, 549 (1957) [Sov. Phys.
JETP \textbf{6}, 429 (1958)]; J. Exptl. Theoret. Phys. {\bf 34}, 247 (1958)
[Sov. Phys. JETP \textbf{7}, 172 (1958)].  

\bibitem{MNS}
Z. Maki, M. Nakagawa, and S. Sakata, Prog. Theor. Phys. \textbf{28}, 870 (1962).


\bibitem{fitvalle}
D.V. Forero et al., Phys. Rev. D90, 093006 (2014).

\bibitem{APL-DRA}
D. C. Rivera-Agudelo and A. P\'erez-Lorenzana, Phys. Rev. D {\bf 92}, 073009
(2015)

\bibitem{mutau}
R.N. Mohapatra, S. Nussinov, \prd {\bf 60}, 013002 (1999); 
C.S. Lam, Phys. Lett. B{\bf 507}, 214 (2001); 
T. Kitabayashi, M. Yasue, \prd {\bf 67}, 015006 (2003); 
W. Grimus, L. Lavoura, Phys. Lett. B{\bf 572}, 189 (2003); 
Y. Koide, \prd {\bf 69}, 093001 (2004).

\bibitem{others}
For an incomplete list, see for instance,
T. Fukuyama, and H. Nishiura, arXiv:hep-ph/9702253;  
P. F. Harrison, W. G. Scott, Phys. Lett.B{\bf 547}, 219 (2002); 
K. S. Babu, R. N. Mohapatra, Phys. Lett. B{\bf 532}, 77 (2002);
T. Ohlsson, G. Seidl, Nucl. Phys. B{\bf 643}, 247(2002); 
R.~N. Mohapatra, J. High Energy Phys. {\bf 10}, 027 (2004);
A. Ghosal, Mod. Phys. Lett. A{\bf 19}, 2579 (2004); 
E. Ma, Phys. Rev. D{\bf 70}, 031901(R) (2004);  
R. N. Mohapatra, S. Nasri, H.-B. Yu, Phys. Lett. B{\bf 636}, 114 (2006); 
A. S. Joshipura, Eur.Phys. J. C {\bf 53}, 77 (2007); 
K. Fuki, M. Yasue, Nucl. Phys. B {\bf 783},31 (2007); 
Riazuddin, Eur. Phys. J. C{\bf 51}, 697 (2007); 
C. Luhn {\it et al.}, Phys. Lett. B{\bf 652}, 27 (2007);
Y. Koide, E. Takasugi, Phys. Rev. D {\bf 77}, 016006 (2008);
M. Honda, M. Tanimoto, Prog. Theor. Phys. {\bf 119}, 583 (2008) ;
H.Ishimori {\it et al.}, Phys. Lett. B{\bf 662}, 178 (2008);
S.-F Ge, H.-J He, F.-R Yin, JCAP {\bf 1005}, 017 (2010);
S.-F. Ge, D. A. Dicus, W. W. Repko, Phys.Lett. B{\bf 702} 220 (2011);
Hong-Jian He, Fu-Rong Yin, Phys. Rev. D {\bf 84}, 033009 (2011); 
S.-F. Ge, D. A. Dicus, W. W. Repko, Phys. Rev. Lett. B{\bf 108}, 041801 (2012);
J. Talbert, J. High Energy Phys. {\bf 12}, 058 (2014); Hong-Jian He, Xun-Jie Xu, Phys. Rev. D {\bf 86}, 111301 (2012); 	


\bibitem{mtseesaw}
J.C. G\'omez-Izquierdo and A. P\'erez-Lorenzana, Phys. Rev. D. \textbf{77} 113015 
(2008).

\bibitem{revmutau}
Z. Z. Xing and Z. H. Zhao, arXiv:1512.04207 [hep-ph]. 

\bibitem{predictingcp}
M Sruthilaya, {\it et al.}, New J. Phys. \textbf{17}, 083028 (2015); 
Y. Shimizu, {\it et al.}, Mod. Phys. Lett. A {\bf 30} 1550002 (2015).


\bibitem{gupta}
S. Gupta, A.S. Joshipura, K.M. Patel, J. High Energy Phys. {\bf 09} (2013) 035.




\end{thebibliography}
\end{document}